\documentclass[preprint,showpacs,showkeys,aps,prb]{revtex4}
\usepackage[dvips]{graphicx}
\begin{document}

\title{
Singly-Thermostated Ergodicity in Gibbs' Canonical Ensemble \\
and the 2016 Ian Snook Prize \\
}

\author{
William Graham Hoover and Carol Griswold Hoover               \\
Ruby Valley Research Institute                  \\
Highway Contract 60, Box 601                    \\
Ruby Valley, Nevada 89833                       \\
}

\date{\today}


\vspace{0.1cm}

\begin{abstract}
For a harmonic oscillator, Nos\'e's single-thermostat approach to simulating Gibbs'
canonical ensemble with dynamics samples only a small fraction of the phase space.
Nos\'e's approach has been improved in a series of three steps: [ 1 ] several
two-thermostat sets of motion equations have been found which cover the complete
phase space in an ergodic fashion; [ 2 ] sets of single-thermostat motion equations,
exerting ``weak control'' over both forces and momenta, have been shown to be ergodic;
and [ 3 ] sets of single-thermostat motion equations exerting weak control over two
velocity moments provide ergodic phase-space sampling for the oscillator and for the
rigid pendulum, but not for the quartic oscillator or for the Mexican Hat potential.
The missing fourth step, motion equations providing ergodic sampling for anharmonic
potentials requires a further advance. The 2016 Ian Snook Prize will be awarded to
the author(s) of the most interesting  original submission addressing the problem of
finding ergodic algorithms for Gibbs' canonical ensemble using a single thermostat.
\end{abstract}

\maketitle

\section{Gibbs' Canonical Ensemble}
From Gibbs' 1902 text {\it Elementary Principles in Statistical Mechanics}, page 183 :
\begin{quotation}
``If a system of a great number of degrees of freedom is microcanonically distributed
in phase, any very small part of it may be regarded as canonically distributed.''
\end{quotation}
Thus J. Willard Gibbs pointed out that the energy states of a ``small'' system weakly
coupled to a larger ``heat reservoir'' with a temperature $T$ have a ``canonical''
distribution :
$$
f(q,p) \propto e^{-{\cal H}(q,p)/kT} \ .
$$
with the Hamiltonian ${\cal H}(q,p)$ that of the small system. Here $(q,p)$ represents
the set of coordinates and momenta of that system.

`` {\it Canonical} '' means simplest or prototypical.  The heat reservoir coupled to
the small system and responsible for the canonical distribution of energies is best
pictured as an ideal-gas thermometer characterized by an unchanging kinetic temperature
$T$ . The reservoir gas consists of many small-mass classical particles engaged in a
chaotic and ergodic state of thermal and mechanical equilibrium with negligible
fluctuations in its temperature and pressure.  Equilibrium within this thermometric
reservoir is maintained by collisions as is described by Boltzmann's equation. His
``H Theorem'' establishes the Maxwell-Boltzmann velocity distribution found in the
gas.  See Steve Brush's 1964 translation of Boltzmann's 1896 text {\it Vorlesungen
\"uber Gastheorie}.

Prior to fast computers texts in statistical mechanics were relatively formal with
very few figures and only a handful of numerical results.  In its more than
700 pages Tolman's 1938 tome {\it The Principles of Statistical Mechanics} includes
only two Figures.  [ The more memorable one, a disk colliding with a triangle,
appears on the cover of the Dover reprint volume. ]  Today the results-oriented
graphics situation is entirely different as a glance inside any recent issue of
{\it Science} confirms.

\section{Nos\'e-Hoover Canonical Dynamics -- Lack of Ergodicity}
In 1984, with the advent of fast computers and packaged computer graphics software
already past, Shuichi Nos\'e set himself the task of generalizing molecular dynamics
to mimic Gibbs' canonical distribution\cite{b1,b2}.  In the end his approach was
revolutionary. It led to a new form of heat reservoir described by a single degree of
freedom with a logarithmic potential, rather than the infinitely-many oscillators or
gas particles discussed in textbooks.  Although the theory underlying Nos\'e's approach
was cumbersome Hoover soon pointed out a useful simplification\cite{b3,b4} : Liouville's
flow equation in the phase space provides a direct proof that the ``Nos\'e-Hoover''
motion equations are consistent with Gibbs' canonical distribution.  Here are the
motion equations for the simplest interesting system, a single one-dimensional
harmonic oscillator :
$$
\dot q =  (p/m) \ ; \ \dot p = -\kappa q - \zeta p \ ; \ \dot \zeta =
[ \ (p^2/mkT) - 1 \ ]/\tau^2 \ .
$$
The ``friction coefficient'' $\zeta$ stabilizes the kinetic energy $(p^2/2m)$ through
integral feedback, extracting or inserting energy as needed to insure a time-averaged
value of precisely $(kT/2)$ .  The parameter $\tau$ is a relaxation time governing
the rate of the thermostat's response to thermal fluctuations.  In what follows we
will set all the parameters and constants $(m,\kappa,k,T,\tau)$ equal to unity,
purely for convenience.  Then the Nos\'e-Hoover equations have the form :
$$
\dot q = p \ ; \ \dot p = -q -\zeta p \ ; \ \dot \zeta = p^2 - 1 \ [ \ {\rm NH} \ ] \ .
$$

Liouville's phase-space flow equation, likewise written here for a single degree of
freedom, is just the usual continuity equation for the three-dimensional flow of a
probability density in the ($q,p,\zeta$) phase space :
$$
\dot f = (\partial f/\partial t) + \dot q(\partial f/\partial q)
+ \dot p(\partial f/\partial p)
+ \dot \zeta(\partial f/\partial \zeta) = -f(\partial \dot q/\partial q)
-f(\partial \dot p/\partial p)-f(\partial \dot \zeta/\partial \zeta) \ .
$$
This approach leads directly to the simple [ NH ] dynamics described above.
It is easy to verify that Gibbs' canonical distribution needs only to be multiplied by
a Gaussian distribution in $\zeta$ in order to satisfy Liouville's equation.
$$
e^{-q^2/2}e^{-p^2/2}e^{-\zeta^2/2} \propto f_{NH} \propto f_Ge^{-\zeta^2/2}
\longrightarrow (\partial f_{NH}/\partial t) \equiv 0 \  .
$$

Hoover emphasized that the simplest thermostated system, a harmonic oscillator,
does {\it not} fill out the entire Gibbs' distribution in $(q,p,\zeta)$ space.  It is not
``ergodic'' and fails to reach all of the oscillator phase space.  In fact,
with {\it all} of the parameters ( mass, force constant, Boltzmann's constant,
temperature, and relaxation time $\tau$  ) set equal to unity only six percent of the
Gaussian distribution is involved in the chaotic sea\cite{b5}.  See {\bf Figure 1} for a
cross section of the Nos\'e-Hoover sea in the $p=0$ plane.  The complexity in the figure,
where the ``holes'' correspond to two-dimensional tori in the three-dimensional
$(q,p,\zeta)$ phase space, is due to the close relationship of the Nos\'e-Hoover
thermostated equations to conventional chaotic Hamiltonian mechanics with its
infinitely-many elliptic and hyperbolic points.

\begin{figure}
\includegraphics[width=4.5in,angle=-90.]{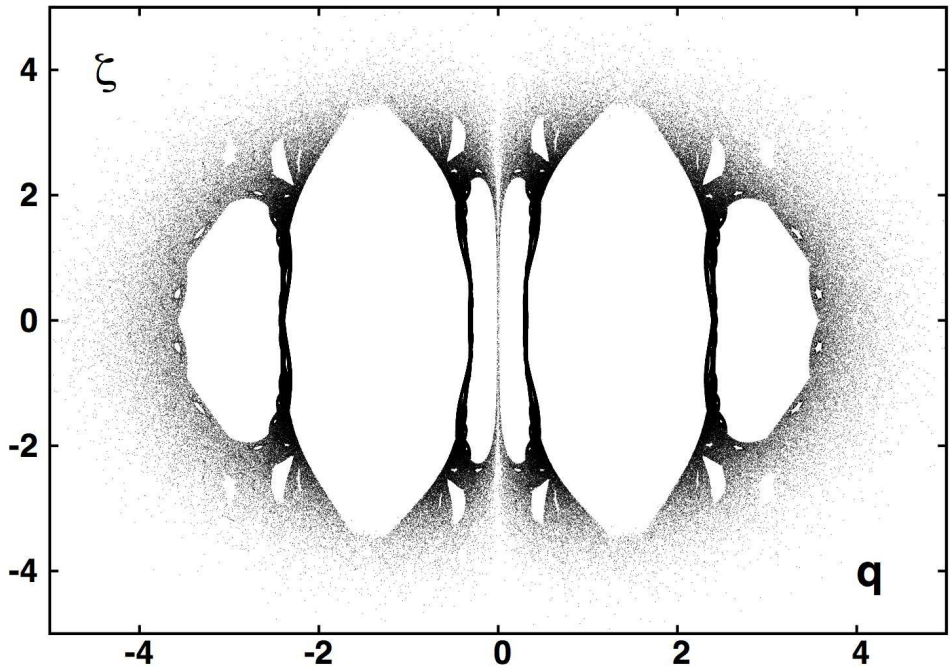}
\caption{
The $p=0$ cross section of the chaotic sea for the Nos\'e-Hoover harmonic oscillator.  
502 924 crossings of the plane are shown.  The fourth-order Runge-Kutta integration
used a timestep $dt = 0.0001$. A point was plotted whenever the product $p_{old}p_{new}$
was negative.
}
\end{figure}

\section{More General Thermostat Ideas}

New varieties of thermostats, some of them Hamiltonian and some not, appeared over the
ensuing 30-year period following Nos\'e's work\cite{b6,b7,b8,b9,b10,b11,b12,b13,b14,b15,
b16,b17,b18}.  This list is by no means complete. 
Though important, simplicity is not the sole motivation for abandoning purely-Hamiltonian
thermostats.  Relatively recently we pointed out that Hamiltonian thermostats are
incapable of generating or absorbing heat flow\cite{b6,b7}.  The close connection
between changing phase volume and entropy production guarantees that Hamiltonian
mechanics is fundamentally inconsistent with irreversible flows.

At equilibrium Bra\'nka, Kowalik, and Wojciechowski\cite{b8} followed Bulgac and
Kusnezov\cite{b9,b10} in emphasizing that {\it cubic} frictional forces, $-\zeta^3p$,
which also follow from a novel Hamiltonian, promote a much better coverage of phase
space, as shown in {\bf Figure 2} . The many small holes in the $p=0$ cross section
show that this approach also lacks ergodicity.

\begin{figure}
\includegraphics[width=4.5in,angle=-90.]{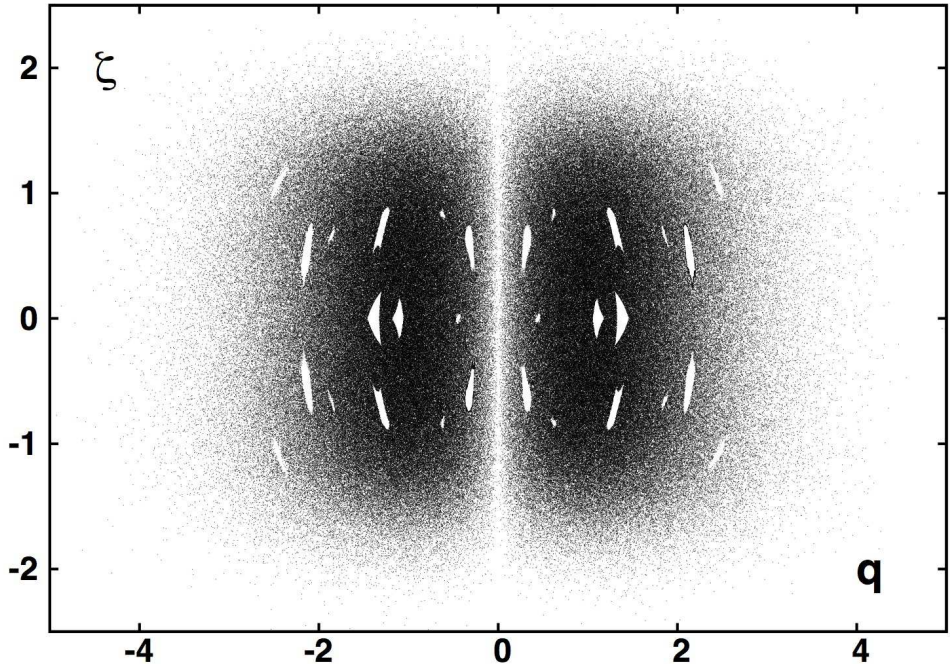}
\caption{
The $p=0$ cross section of the chaotic sea for an oscillator governed by Bra\'nka, Kowalik,
and Wojciechowski's choice of the motion equation, $\ddot q = \dot p = -q -\zeta^3p \ ; \
\dot \zeta = p^2 - 1$ .
20 billion timesteps, with $dt = 0.0001$, resulted in 636 590 crossings of the $p=0$ section,
using the integration procedure of Figure 1.
}
\end{figure}

\subsection{Joint Control of Two Velocity Moments}

Attempts to improve upon this situation led to a large literature with the most useful
contributions applying thermostating ideas with two or more thermostat variables\cite{b9,b10}.
An example, applied to the harmonic oscillator, was tested by Hoover and Holian\cite{b11}
and found to provide all of Gibbs' distribution :
$$
\dot q = p \ ; \ \dot p = -q - \zeta p - \xi p^3\ ; \
\dot \zeta = p^2 - 1 \ ; \ \dot \xi = p^4 - 3p^2 \ {\rm [ \ HH \ ]}
$$
The two thermostat variables $(\zeta,\xi)$ together guarantee that both the second
and the fourth moments of the velocity distribution have their Maxwell-Boltzmann values
[ 1 and 3 ] .  Notice that two-dimensional cross sections like those in the Figures are no
longer useful diagnostics for ergodicity once the phase-space dimensionality exceeds three.

\subsection{Joint Control of Coordinates and Velocities}

In 2014 Patra and Bhattacharya\cite{b12} suggested thermostating both the coordinates and the
momenta :
$$
\dot q = p - \xi q \ ; \ \dot p = -q - \zeta p  \ ; \
\dot \zeta = p^2 - 1 \ ; \ \dot \xi = q^2 - 1 \ {\rm [ \ SEPB \ ]} \ .
$$
an approach already tried by Sergi and Ezra in 2001\cite{b13}.

A slight variation of the Sergi-Ezra-Patra-Bhattacharya thermostat takes into account Bulgac
and Kusnezov's observation that cubic terms favor ergodicity :
$$
\dot q = p - \xi^3 q \ ; \ \dot p = -q - \zeta p  \ ; \
\dot \zeta = p^2 - 1 \ ; \ \dot \xi = q^2 - 1 \ {\rm [ \ PB_{var} \ ]} \ . 
$$
These last two-thermostat equations appear to be a good candidate for ergodicity, reproducing
the second and fourth moments of $(q,p,\zeta,\xi)$ within a fraction of a percent.  We have not
carried out the thorough investigation that would be required to establish their ergodicity
as the single-thermostat models are not only simpler but also much more easily diagnosed 
because their sections are two-dimensional rather than three-dimensional.
 
\section{Single-Thermostat Ergodicity}

Combining the ideas of ``weak control'' and the successful simultaneous thermostating of
coordinates and momenta\cite{b14} led to further trials attempting the weak control of two
different kinetic-energy moments\cite{b15}.  One choice out of the hundreds investigated
turned out to be successful for the harmonic oscillator :
$$
\dot q = p \ ; \ \dot p = - q -\zeta( 0.05p + 0.32p^3) \ ; \
\dot \zeta = 0.05(p^2 - 1) + 0.32(p^4 - 3p^2) \ [ \ {\rm ``0532 \ Model''} \ ] \ .
$$
These three oscillator equations passed all of the following tests for ergodicity :

\noindent
[ 1 ] The moments $\langle \ p^2 \ \rangle = 1 \ ; \ \langle \ p^4\ \rangle = 3 \ ; \
\langle \ p^6 \ \rangle = 15 $ were confirmed.

\noindent
[ 2 ] The independence of the largest Lyapunov exponent to the initial conditions indicated
the absence of the toroidal solutions.

\noindent
[ 3 ] The separation of two nearby trajectories had an average value of 6 :\\
$\langle \ (q_1-q_2)^2 + (p_1-p_2)^2 + (\zeta_1-\zeta_2)^2 \ \rangle = 2 + 2 + 2 = 6 $ .

\noindent
[ 4 ] The times spent at positive and negative values of $\{ \ q,p,\zeta \ \}$ were close to
 equal.

\noindent
[ 5 ] The times spent in regions with each of the 3! orderings of the three dependent variables
were equal for long times.

These five criteria were useful tools for confirming erogidicity.  Evidently weak control is the
key to efficient ergodic thermostating of oscillator problems.

\begin{figure}
\includegraphics[width=4.5in,angle=-90.]{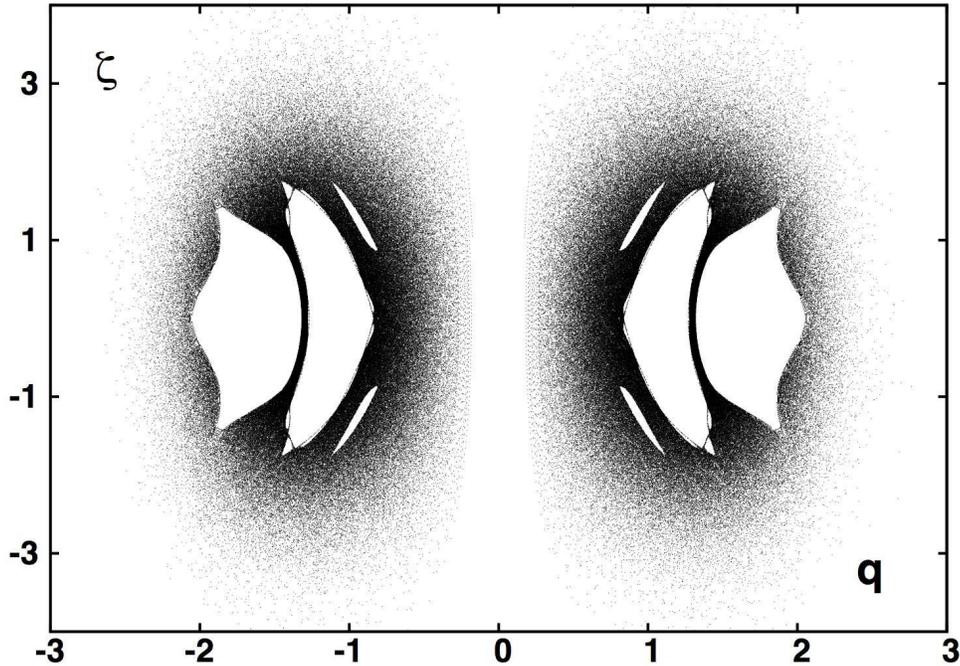}
\caption{
$p=0$ cross section for a singly-thermostated quartic oscillator, with motion equations $ \ddot q
= \dot p = -q^3 -\zeta p^3 \ ; \ \dot \zeta = p^4 - 3p^2$ .  Runge-Kutta integration as in Figures
1 and 2 with 503 709 crossings of the $p=0$ plane.  Several hundred singly-thermostated attempts
failed to obtain canonical ergodicity for the quartic oscillator.
}
\end{figure}

\section{A Fly in the Ointment, the Quartic Potential}

The success in thermostating the harmonic oscillator led to like results for the simple pendulum
but {\it not} for the quartic potential\cite{b15}.  See {\bf Figure 3}. This somewhat surprising
setback motivates the need for more work and is the subject of the Ian Snook Prize for 2016.  This
Prize will be awarded to the author(s) of the most interesting original work exploring the
ergodicity of single-thermostated statistical-mechanical systems.  The systems
are not at all limited to the examples of the quartic oscillator and the Mexican Hat potential 
but are left to the imagination and creativity of those entering the competition.

\begin{figure}
\includegraphics[width=2.0in,angle=-0.]{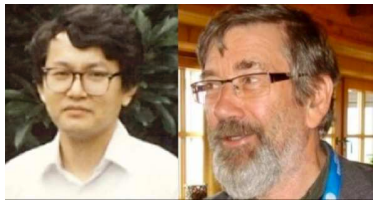}
\caption{
Shuichi Nos\'e ( 1951-2005 ) and Ian Snook ( 1945-2013 )
}
\end{figure}
 
\section{Conclusions -- Ian Snook Prize for 2016}

It is our intention to reward the most interesting and convincing entry submitted for publication
to Computational Methods in Science and Technology ( www.cmst.eu ) prior to 31 January 2017. The
2016 Ian Snook prize of \$500 dollars will be presented to the winner in early 2017.  An Additional
Prize of the same amount will likewise be presented by the Institute of Bioorganic Chemistry of
the Polish Academy of Sciences ( Poznan Supercomputing and Networking Center ). We are grateful
for your contributions.  This work is dedicated to the memories of our colleagues, Ian Snook
( 1945-2013 ) and Shuichi Nos\'e ( 1951-2005 ), shown in {\bf Figure 4} .
\pagebreak

\end{document}